# Synthetic Expertise

Ron Fulbright[1] and Grover Walters[2]

University of South Carolina Upstate
800 University Way, Spartanburg, SC 29303
[1]`rfulbright@uscupstate.edu`
[2]`gwalters@uscupstate.edu`

**Abstract.** We will soon be surrounded by artificial systems capable of cognitive performance rivaling or exceeding a human expert in specific domains of discourse. However, these "cogs" need not be capable of full general artificial intelligence nor able to function in a stand-alone manner. Instead, cogs and humans will work together in collaboration each compensating for the weaknesses of the other and together achieve synthetic expertise as an ensemble. This paper reviews the nature of expertise, the Expertise Level to describe the skills required of an expert, and knowledge stores required by an expert. By collaboration, cogs augment human cognitive ability in a human/cog ensemble. This paper introduces six Levels of Cognitive Augmentation to describe the balance of cognitive processing in the human/cog ensemble. Because these cogs will be available to the mass market via common devices and inexpensive applications, they will lead to the Democratization of Expertise and a new cognitive systems era promising to change how we live, work, and play. The future will belong to those best able to communicate, coordinate, and collaborate with cognitive systems.

## 1 Introduction

The idea of enhancing human cognitive ability with artificial systems is not new. In the 1640s, mathematician Blaise Pascal created a mechanical calculator [1]. Yet, the Pascaline, and the abacus thousands of years before that, were just mechanical aids executing basic arithmetic operations. The human does all the real thinking. In the 1840s, Ada Lovelace envisioned artificial systems based on Babbage's machines assisting humans in musical composition [2][3]. In the 1940s, Vannevar Bush envisioned the Memex and discussed how employing associative linking could enhance a human's ability to store and retrieve information [4]. The Memex made the human more efficient but did not actually do any of the thinking on its own. In the 1950s, Ross Ashby coined the term *intelligence amplification* maintaining human intelligence could be synthetically enhanced by increasing the human's ability to make appropriate selections on a persistent basis [5]. But again, the human does all of the thinking. The synthetic aids just make the human more efficient. In the 1960s, J.C.R. Licklider and Douglas Engelbart envisioned human/computer symbiosis—humans and artificial systems working together in co-dependent fashion to achieve performance greater than either could by working alone [17][18].

Over thirty years ago, Apple, Inc. envisioned an intelligent assistant called the *Knowledge Navigator* [6]. The Knowledge Navigator was an artificial executive assistant capable of natural language understanding, independent knowledge gathering and processing, and high-level reasoning and task execution. The Knowledge Navigator concept was well ahead of its time and not taken seriously. However, some of its features are seen in current voice-controlled "digital assistants" such as Siri, Cortana, and the Amazon Echo (Alexa). Interestingly, the Knowledge Navigator was envisioned as a collaborator rather than a stand-alone artificial intelligence, a feature we argue is critical.

In 2011, a cognitive computing system built by IBM, called Watson, defeated two of the most successful human Jeopardy champions of all time [7]. Watson received clues in natural language and gave answers in natural spoken language. Watson's answers were the result of searching and deeply reasoning about millions of pieces of information and aggregation of partial results with confidence ratios. Watson was programmed to *learn* how to play Jeopardy, which it did in many training games with live human players before the match [8][9]. Watson *practiced* and achieved expert-level performance within the narrow domain of playing Jeopardy. Watson represents a new kind of computer system called *cognitive systems* [10][3]. IBM has been commercializing Watson technology ever since.

In 2016, Google's AlphaGo defeated the reigning world champion in Go, a game vastly more complex than Chess [13][14]. In 2017, a version called AlphaGo Zero learned how to play Go by playing games with itself and not relying on any data from human games [15]. AlphaGo Zero exceeded the capabilities of AlphaGo in only three days. Also in 2017, a generalized version of the learning algorithm called AlphaZero was developed capable of learning any game. While Watson required many person-years of engineering effort to program and teach the craft of Jeopardy, AlphaZero achieved expert-level performance in the games of Chess, Go, and Shogi after only a few hours of unsupervised self-training [16].

These recent achievements herald a new type of artificial entity, one able to achieve, in a short amount of time, expert-level performance in a domain without special knowledge engineering or human input. Beyond playing games, artificial systems are now better at predicting mortality than human doctors [40], detecting signs of child depression through speech [41], detecting lung cancer in X-Rays [42, 43]. Systems can even find discoveries in old scientific papers missed by humans [44].

In 2014, IBM released a video showing two humans interacting with and collaborating with an artificial assistant based on Watson technology [11]. IBM envisions systems acting as partners and collaborators with humans. The similarity between the IBM Watson video and the Knowledge Navigator is striking. John Kelly, Senior Vice President and Director of Research at IBM describes the coming revolution in cognitive augmentation as follows [12]:

> "The goal isn't to replace human thinking with machine thinking. Rather humans and machines will collaborate to produce better results—each bringing their own superior skills to the partnership."

We believe the ability for systems to learn on their own how to achieve expert-level performance combined with cognitive system technology will lead to a multitude of mass-market apps and intelligent devices able to perform high-level cognitive processing. Millions of humans around the world will work daily with and collaborate with these systems we call *cogs*. This future will belong to those of us better able to collaborate with these systems to achieve expert-level performance in multiple domains—*synthetic expertise*.

## 2 Literature

**2.1 Human/Computer Symbiosis**
Engelbart and Licklider envisioned human/computer symbiosis in the 1960s. Licklider imagined humans and computers becoming mutually interdependent, each complementing the other [17]. However, Licklider envisioned the artificial aids merely assisting with the preparation work leading up to the actual thinking which would be done by the human. In 1962, Engelbart developed the famous H-LAM/T framework modeling an augmented human as part of a system consisting of: the human, language (concepts, symbols, representations), artifacts (physical objects), methodologies (procedures, know-how), and training [18]. As shown in Fig. 1, Engelbart's framework envisions a human interacting, and working together on a task, with an artificial entity To perform the task, the system executes a series of processes, some performed by the human (explicit-human processes), others performed by artificial means (explicit-artifact processes), and still others performed by a combination of human and machine (composite processes). Engelbart's artifacts were never envisioned to do any of the high-level thinking. We feel as though the *artifacts* themselves are about to change. Recent advances in machine learning and artificial intelligence research indicates the artifacts are quickly becoming able to perform human-like cognitive processing.

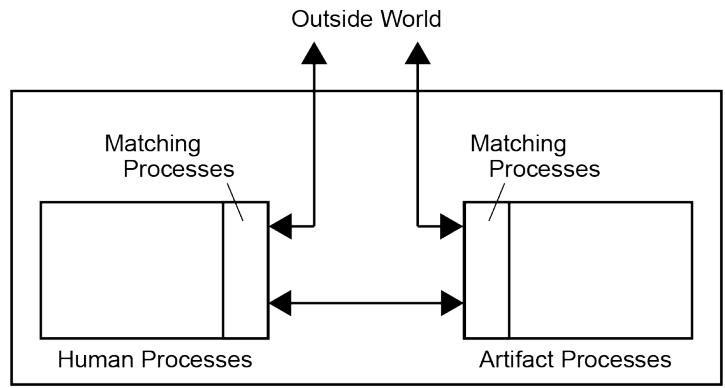

**Fig. 1**. Engelbart's H-LAM/T Framework

**2.2 The Nature of Expertise**
What it means to be an expert has been debated for decades. Traditional definitions of expertise rely on *knowledge* (the know-that's) and *skills* (the know-how's). Gobet offers a more general definition maintaining an expert *"obtains results vastly superior to those obtained by the majority of the population"* [34]. De Groot established the importance of *perception* in expertise in that an expert perceives the most important aspects of a situation faster than novices [45]. Formally, experts are goal-driven intelligent agents where set of goals, *G,* and a set of utility values, *U,* drive the expert's perceiving, reasoning and acting over time [23][24]. Intelligent agents perceive a subset, *T*, of possible states, *S*, of the environment and perform actions from set of actions, *A,* to effect changes on the environment.

In their influential study of experts, Chase and Simon found experts compile a large amount of domain-specific knowledge from years of experience—on the order of 50,000 pieces [21]. Steels later describes this as *deep domain knowledge* and identified: *problem-solving methods*, and *task models* as needed by an expert [22]. Problem-solving methods are to solve a problem and a task model is knowledge about how to do something. An expert must know both generic and domain-specific problem-solving methods and tasks. In a new situation, an expert can perceive the most important features quicker than a novice. Experts then match the current situation to their enormous store of deep domain knowledge and efficiently extract knowledge and potential solutions from memory. Furthermore, an expert applies this greater knowledge to the situation more efficiently and quicker than a novice making experts superior problem solvers.

Fulbright extended the description of experts by defining the fundamental skills of an expert based on the skills identified in Bloom's Taxonomy [46]. These skills, described at the Expertise Level, and the knowledge stores, described at the Knowledge Level, form Fulbright's Model of Expertise shown in Fig. 2.

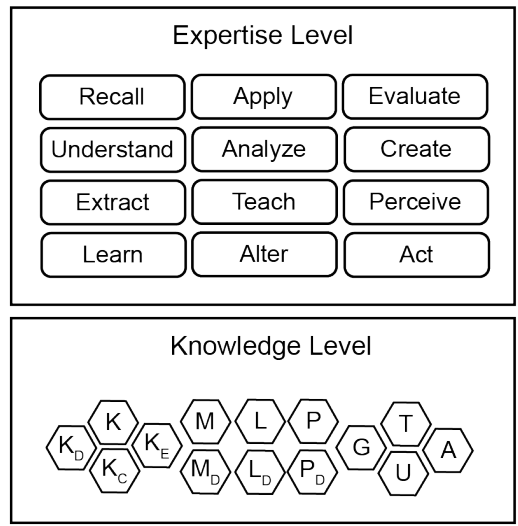

*Knowledge*
**K**     declarative knowledge statements
**$K_D$**   domain-specific knowledge
**$K_C$**   common-sense knowledge
**$K_E$**   episodic knowledge       **G**  goals to achieve
**M/$M_D$** world models           **U**  utility values
**L/$L_D$** task models            **T**  perceivable states
**P/$P_D$** problem-solving models   **A**  actions

*Skills*
**Perceive**   sense/interpret the environment
**Act**          perform action affecting environment
**Recall**      remember; store/retrieve knowledge
**Understand** classify, categorize, discuss, explain, identify
**Apply**       implement, solve, use knowledge
**Analyze**    compare, contrast, experiment
**Evaluate**   appraise, judge, value, critique
**Create**      design, construct, develop, synthesize
**Extract**     match/retrieve deep knowledge
**Learn**       modify existing knowledge
**Teach**      convey knowlege/skills to others
**Alter**       modify goals

**Fig. 2**. Model of Expertise

Cognitive scientists have studied and modeled human cognition for decades. The most successful cognitive architecture to date, begun by pioneer Allen Newell and now led by John Laird, is the Soar architecture [25]. Fulbright recently applied the Model of Expertise shown in Fig. 2 to the Soar architecture as shown in Fig. 3 [46].

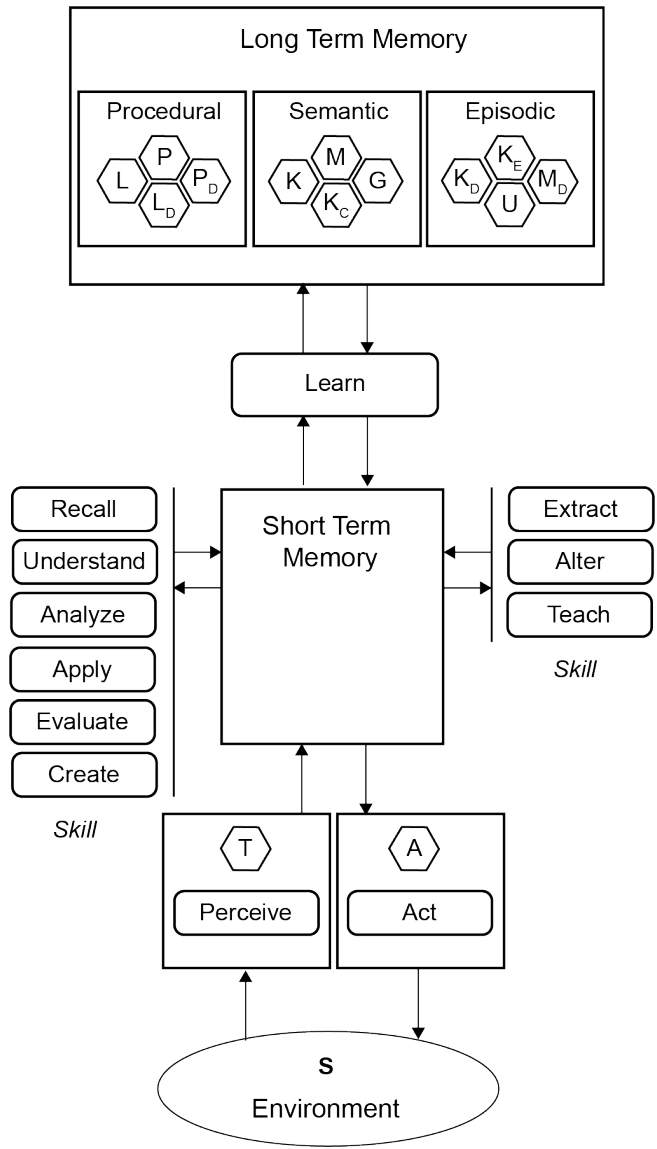

**Fig. 3**. Soar Model of Expertise

**2.3 Human Cognitive Augmentation**

Fulbright described a human/cog ensemble, much like Engelbart's HLAM/T framework, where the human does some cognitive processing and the cog does some cognitive processing [28]. With $W$ being a measure of cognitive processing, comparing the amount of cognitive processing done by each component yields a metric called the *augmentation factor*, $A^+$

$$A^+ = \frac{W_{cog}}{W_{hum}} \qquad (1)$$

If the human does more cognitive processing than the cog, $A^+ < 1$; but when the cog starts performing more cognitive processing than the human, $A^+ > 1$ and increases without bound as the capability of cogs grows. At some point in the future, human cognitive processing may become vanishingly small relative to the cog. When cogs develop to the point of being truly *artificial experts,* able to perform at the expert level without human contribution, the human component in Eq. (1) falls to zero and the idea of being cognitively augmented will be a senseless quantity to measure. Until that time though, the world will belong to those humans best able to partner with and collaborate with cogs.

These humans will be cognitively augmented. A human working alone performs $W_{hum}$ amount of cognitive processing. However, a human working with one or more cogs performs an increased amount of cognitive processing [29][30]

$$W^* = W_{hum} + \sum W_{cog}^i \qquad (2)$$

where *i* is the number of cogs in the collaboration and $W^*$ is the total cognitive processing done by the ensemble. To an observer outside of the ensemble, it appears the human is performing at a much higher level than expected. When $W^*$ reaches or exceeds the level of an expert in the domain of discourse, the human/cog ensemble will have achieved *synthetic expertise* as shown in Equation (3).

$$W^* \geq W_{expert} \qquad (3)$$

Fulbright described the measurement of the augmentation of two specific cognitive capabilities, *cognitive accuracy* and *cognitive precision* [30]. Given a problem to solve, cognitive accuracy involves the ability to synthesize the best solution and cognitive precision involves being able to synthesize nothing but the best solution. In a case study, with the cog component being simulated by expert "suggestions," cognitive accuracy of the human was improved by 74% and cognitive precision was improved by 28%.

## 3  Synthetic Expertise

Biological systems capable of performing all skills and acquiring/possessing all knowledge in the Model of Expertise shown in Fig. 2 are *human experts*. Non-biological systems capable of the same are called *artificial experts.* We envision a future in which artificial experts achieve or exceed the performance of human experts in virtually every domain. However, we think it will be some time before fully autonomous artificial experts exist. In the immediate future, humans and artificial systems will work together to achieve expertise as an ensemble—*synthetic expertise.* We choose the word "synthetic" rather than "artificial" because the word artificial carries a connotation of not being real. We feel as though the cognitive processing performed by cogs and the ensemble is real even though it may be very different from human cognitive processing.

We call the artificial collaborators *cogs.* Cogs are intelligent agents—entities able to rationally act toward achieving a goal [24]. However, the term *intelligent agent* refers to a wide range of systems, from very simple systems such as a thermostat in your home to very complex systems, such as artificially intelligent experts. In the study of synthetic expertise, the term *cog* is defined as:

> ***cog:*** an intelligent agent, device, or algorithm able to perform, mimic, or replace one or more cognitive processes performed by a human or a cognitive process needed to achieve a goal.

It is important to note cogs can be artificially intelligent but do not necessarily have to be. However, cogs are expected to be relatively complex because, for synthetic expertise, cogs should perform part of, or all of, at least one of the fundamental skills identified in the Model of Expertise shown in Fig. 2: *recall, apply, understand, evaluate, analyze, extract, alter, learn, teach, perceive, act,* and *create.* Cogs also need not be terribly broad in scope nor deep in performance. They can be narrow and shallow agents. Cogs also need not fully implement a cognitive process, but instead may perform only a portion of a cognitive process with the human performing the remainder of that process. Computers have tremendous advantage over humans in some endeavors such as number crunching, speed of operations, and data storage. Some cogs will leverage their advantage in these kinds of functions in support of a jointly-performed cognitive process. Cognitive processing in. a human/cog ensemble will therefore be a combination of biological cognitive processing and non-biological cognitive processing.

Indeed, we see this already beginning to happen both at the professional level and at the personal level. Every day, millions of people issue voice commands to virtual assistants like Apple's Siri, Microsoft's Cortana, Google Assistant, and Amazon's Alexa and a host of applications on computers and handheld electronic devices. These assistants can understand spoken natural language commands and reply by spoken natural language. In the professional world, professionals such as doctors are using cognitive systems to diagnose malignant tumors and bankers are using cognitive systems to analyze risk profiles. In some domains, the cogs already outperform humans.

At the most basic level we envision a human interacting with a cog as shown in Fig. 4. The human and cog ensemble form an Engelbart-style system with the human component performing some of the cognitive work, $W_H$ and the cog performing some of the cognitive work, $W_C$. Information flows into the ensemble and out of the ensemble, $S_{in}$ and $S_{out}$ respectively, and the total cognitive work performed by the ensemble is $W^*$. The difference between Fig. 4 and Engelbart's HLAM/T framework is the cogs are capable of high-level human-like cognitive processing and act as peer collaborators working with humans rather than mere tools. As cogs become more advanced, the human/cog collaboration will become more collegial in nature.

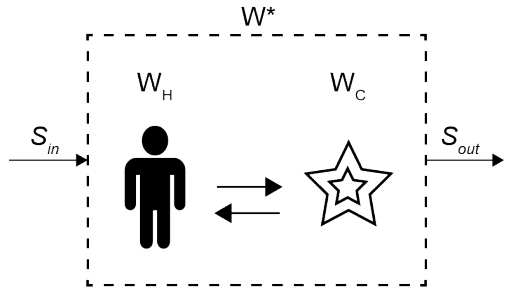

**Fig. 4**. Human/Cog Collaboration

Recalling Equation (3), when the cognitive performance of the ensemble reaches or exceeds that of an expert, the ensemble has achieved *synthetic expertise.* While one day, a single cog may be developed capable of expert performance in any domain, for the foreseeable future, the domain of discourse of the human/cog ensemble will be limited. A cog may help raise a person's performance in only one domain of discourse, or even a subset of a particular domain. Therefore, we expect the near future to see humans employing a number of different cogs, each with different capabilities. In fact, we see this today as well. People employ a number of different apps and devices throughout the day. The difference in the cognitive systems future will be the apps and devices will be capable of high-level cognitive processing. Furthermore, we certainly expect more than one human to be involved in come collaborations forming a virtual team where the cogs will be seen as just another team member.

In this future, communication and collaboration within the ensemble/virtual team is key and is a fertile area for future research. Humans will certainly converse with other humans (human/human). Likewise, cogs will converse with other cogs (cog/cog) and humans will converse with cogs (human/cog). The dynamics of each of these three realms of communication and collaboration are quite different and should be explored in future research. In fact, research is already underway. The fields of human/computer interaction (HCI), human/autonomy teaming (HAT), and augmented cognition (AugCog) are currently quite active. The fields of distributed artificial intelligence (DAI), multiagent systems, negotiation, planning, and communication, and computer-supported cooperative work (CSCW) are older fields of study but are quite relevant. Fields such as human-centered design, augmented reality (AR), virtual reality (VR), enhanced reality (ER), and brain-computer interfaces (BCI) lead in promising directions. Information design (ID), user experience design (UX), and information architecture (IA) have important contributions.

### 3.1 The Human/Cog Ensemble

In the cog future, the *collaborate* skill is critical for both humans and cogs. Because the human and the cog are physically independent agents, both must *perceive, act,* and *collaborate*. Adding *collaborate* to the skills and knowledge stores identified in the Model of Expertise (Fig. 2) yields the depiction of synthetic expertise shown in Fig. 5. The human/cog ensemble must perform all skills and maintain all knowledge stores. Skills are performed solely by the human (corresponding to Engelbart's human-explicit processes), solely by the cog (corresponding to Engelbart's artifact-explicit processes), or by a combination of human and cog effort (Engelbart's composite processes). To the outside world, it does not matter which entity performs a skill as long as the skills are performed by the ensemble.

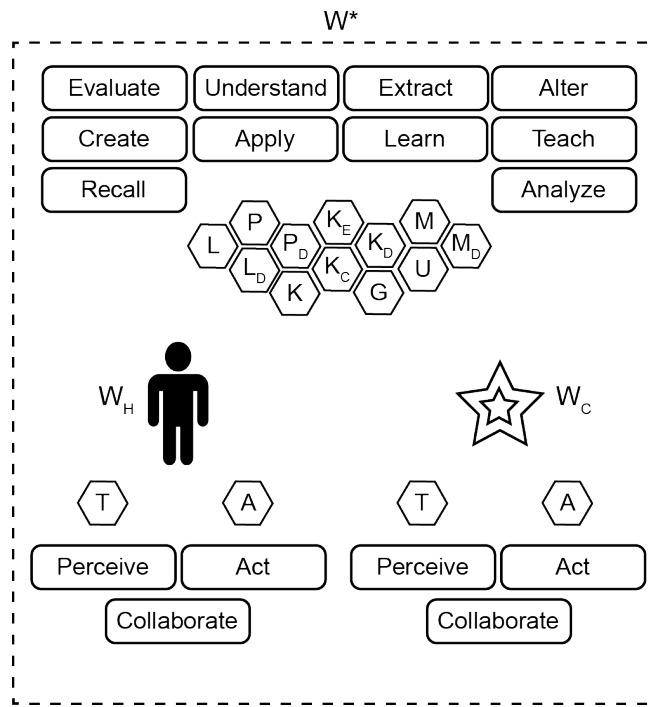

**Fig. 5**. Synthetic Expertise

For the immediate future, cogs will perform lower-order skills and humans will perform higher-order skills. As an example, consider the situation with today's virtual assistants, like Siri, as shown in Fig. 6. Assume a person asks Siri what time it is while performing a task. The human performs an action (*act*) by speaking the command "Siri, what time is it?"

Through the smartphone's microphone, Siri *perceives* the spoken command, and *analyzes* it to *understand* the user is asking for the time (even though this is a rudimentary form of understanding). Siri then *recalls* the time from the internal clock on smartphone, formulates a spoken response, and articulates the reply back to the user (*act*). In this situation, the cog is not doing a large amount of cognitive processing. The human is doing most of the thinking as represented by more of the fundamental skills being shown on the human side. $A^+ < 1$ in this situation. However, as cognitive systems evolve, they will be able to perform more of the higher-order fundamental skills themselves.

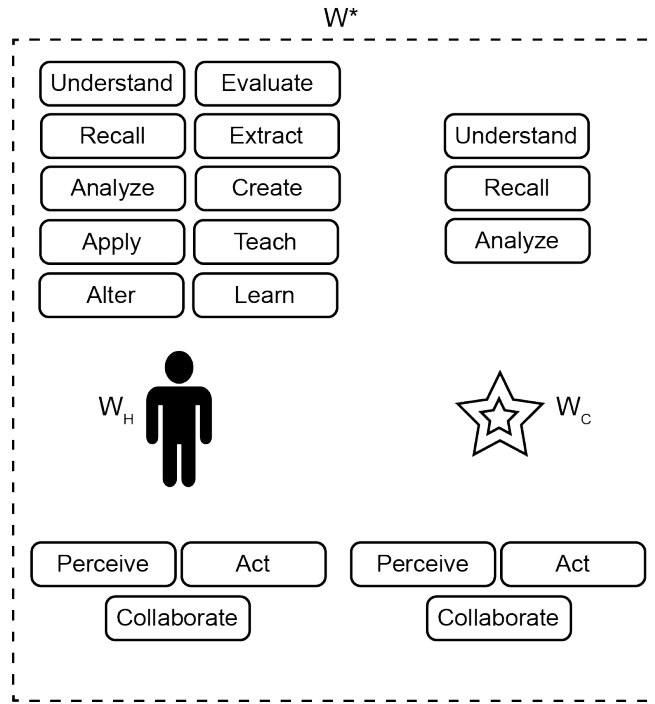

**Fig. 6**. Human-Intensive Expertise

### 3.2 Cognitive Augmentation
In the coming cognitive systems era, cognitive processing in human/cog ensembles will be a mixture of human and cog processing resulting in the augmentation of the human's cognitive processing. It will be many years before fully artificial intelligences become available to the mass market. In the meantime, there will be human/cog ensembles achieving varying amounts of cognitive augmentation. Here, we define the Levels of Cognitive Augmentation ranging from no augmentation at all (all human thinking) to fully artificial intelligence (no human thinking) as shown in Fig. 7.

Until now, computers and software humans use represent Level 1 cognitive augmentation (assistive tools). Recent advances in deep learning and unsupervised learning have produced Level 2 cognitive augmentation. But as the abilities of cogs improves, we will see Level 3 and Level 4 cognitive augmentation leading eventually to fully artificial intelligence, Level 5, in which no human cognitive processing will be required.

**Level 0**: No Augmentation
    human performs all cogntiive processing

**Level 1**: Assistive Tools
    abacus, calculators, software, etc.

**Level 2**: Low-Level Cognition
    pattern recognition, classification, speech
    human makes all high-level decisions

**Level 3**: High-Level Cognition
    concept understanding, critique,
    conversational natural language

**Level 4**: Creative Autonomy
    human-inspired, unsupervised synthesis

**Level 5**: Artificial Intelligence
    no human cognitive processing

**Fig. 7**. Levels of Cognitive Augmentation

The promise of cogs, intelligent agents, cognitive systems, and artificial intelligence in general, is superior performance. If ***P*** is a measure of human performance working alone and ***P*** * is a measure of human/cog performance, then we expect

$$P^* > P \tag{4}$$

so we can calculate the percentage change realized by the human working with a cog

$$\Delta P = \frac{P^* - P}{P}. \tag{5}$$

How does one measure performance in a particular domain of discourse? This may vary widely from domain to domain but in general, we may seek to reduce commonly measured quantities such as *time*, *effort*, and *cost*. Or we may seek to increase quantities such as:

- Quality
- Revenue
- Efficiency
- Number of Transactions
- Number of Actions Completed
- Number of Customers Serviced
- Level of Cognition Achieved

As an example, consider a deep-learning algorithm able to detect lung cancers better than human doctors [42]. The rate of false positives and false negatives by human evaluation of low-dose computed tomography (LDCT) scans delay treatment of lung cancers until the cancer has reached an advanced stage. However, the algorithm outperforms humans in recognizing problem areas reducing false positives by 11% and false negatives by 5%. Therefore, the human/cog ensemble achieves better performance by a measurable extent. Another way of putting this is by working with the cog, the doctor's performance in enhanced.

In dermatology, Google's Inception v4 (a convolutional neural network) was trained and validated using dermoscopic images and corresponding diagnoses of melanoma [47]. Performance of this cog against 58 human dermatologists was measured using a 100-image testbed. Measured was the sensitivity (the proportion of people with the disease with a positive result), the specificity (the proportion of people without the disease with a negative result), and the ROC AUC (a performance measurement for classification problem at various thresholds settings). Results are shown in Fig. 8. The cog outperformed the group of human dermatologists by significant percentages suggesting in the future, the human dermatologists would improve their performance by working with this cog.

|              | Human | Cog   | Improvement |
|--------------|-------|-------|-------------|
| Sensitivity: | 86.6% | 95.0% | +9.7%       |
| Specificity: | 71.3% | 82.5% | +15.7%      |
| ROC AUC:     | 0.79  | 0.86  | +8.9%       |

**Fig. 8**. Human vs. Cog in Lesion Classification

In the field of diabetic retinopathy, a study evaluated the diagnostic performance of an autonomous artificial intelligence system, a cog, for the automated detection of diabetic retinopathy (DR) and Diabetic Macular Edema (DME) [48]. The cog exceeded all pre-specified superiority goals as shown in Fig. 9.

|              | Goal    | Cog   | Improvement |
|--------------|---------|-------|-------------|
| Sensitivity: | >85.0%  | 87.2% | +2.6%       |
| Specificity: | >82.5%  | 90.7% | +9.9%       |

**Fig. 9**. Cog Performance in Diabetic Retinopathy

This begs an important question. Doctors use other artificial devices to perform their craft. Thermometers and stethoscopes enhance a doctor's performance. Why are cogs different? The answer is yes, these tools enhance human performance. Humans have been making and using tools for millennia and indeed this is one differentiating characteristic of humans. Engelbart and Licklider's vision of "human augmentation" in the 1960s was for computers to be tools making humans better and more efficient at thinking and problem solving. Yet, they envisioned the human as doing most of the thinking. We are now beginning to see cognitive systems able to do more than a mere tool, they are able to perform some of the high-level thinking on their own. Today, some of the highest-level skills identified in the Model of Expertise (Fig. 2) such as *understand* and *evaluate* are beyond current cog technology, but the ability of cognitive systems is gaining rapidly.

Research areas such as task learning, problem-solving method learning, goal assessment, strategic planning, common sense knowledge learning, generalization and specification are all critical areas of future research.

### 3.3 Knowledge of the Ensemble

In a human/cog ensemble, the human will possess some of the knowledge stores identified in the Model of Expertise and the cog will possess some of the knowledge stores. The knowledge of the ensemble should be viewed as a combination of human-maintained knowledge and cog-maintained knowledge. In most cases, we expect both the human and cog to possess and maintain their own versions of the knowledge stores and communicate contents to each other when necessary. However, we recognize existing and future technology able to combine these stores by connecting the human mind directly with a computer. With technology like this, a knowledge store could be shared directly by human and cog without needing communication.

Until such technology becomes available, cogs have a unique and important advantage over humans. Cogs can simply download knowledge from an external source—even another cog. Currently, it is not possible to simply dump information directly into a human brain. However, cogs can simply transmit knowledge directly from an external source, such as the Internet. With global communication via the Internet, cogs will have near instantaneous access to knowledge far beyond its own and be able to obtain this remote knowledge with minimal effort.

Fig. 10 depicts a cog, involved in a local human/cog ensemble, sending and receiving knowledge to and from remote stores via the Internet. The figure shows domain-specific domain knowledge ($K_D$) and domain-specific problem-solving knowledge ($P_D$) being obtained from two different remote sources. As far as the local ensemble is concerned, once downloaded, the knowledge stores obtained remotely are no different from locally-produced knowledge. It is as if the ensemble had always been in possession of this knowledge. Any knowledge store can be imported partially or entirely from a remote source. We have described the human/cog ensemble as a local entity but in reality, with pervasive Internet connectivity, a human/cog ensemble is a combination of local knowledge and all other available knowledge. In the cog era, instead of benefitting from one cog, humans will actually be benefitting from millions of cogs. The local/remote line will tend to blur and this vast artificial knowledge will just be assumed to be available "in the cloud" anytime we want it, much how we view Internet-based services today.

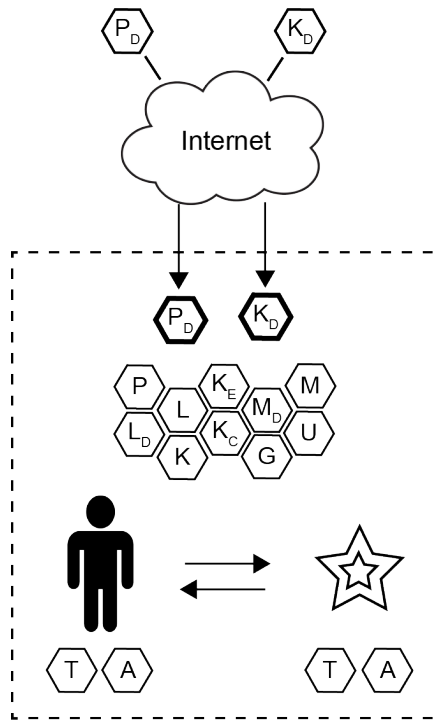

**Fig. 10**. Remote Knowledge Stores

This makes possible one of the most exciting features of the coming cog era. Once a cog learns or synthesizes something, any other cog in existence then or in the future can obtain the knowledge. Millions of cogs interacting with users on a daily basis, continually learning and communicating knowledge to all other cogs, will lead to exponential knowledge creation and evolution of cog capabilities. We expect cogs and cog-based knowledge to evolve very quickly. Fields such as knowledge representation, knowledge management, semantic Web, software agents, distributed artificial intelligence, and ontologies are all important future research areas along with automated knowledge discovery.

### 3.4 Composite Skills

The Model of Expertise shown in Fig. 2, identifies fundamental skills necessary for expertise: *recall, apply, understand, evaluate, analyze, extract, alter, learn, teach, perceive, act,* and *create.* However, one quickly recognizes this is not a complete list of everything an expert does. Experts are certainly expected to do things like: summarize, conceptualize, theorize, classify, categorize, visualize, predict, forecast, define, explain, suggest, compare, assess, prioritize, organize, analyze, recommend, motivate, inspire, etc.

Many of these are included within the scope of the fundamental skills identified and defined in Fig. 2. For example, classify and categorize fall under the *understand* fundamental skill. Others, are higher-order composite skills made up by a combination of fundamental skills. For example, predicting and forecasting involve a combination of the *understand*, *apply*, *create*, and *evaluate* fundamentals. In another example, explaining is a combination of *understanding* and *analyzing*.

A promising area for future research is to define constructions of fundamental skills for higher-order skills. Doing so would make it possible to compare and contrast future implementations of the higher-order skills and possibly lead to establishing effective metrics for the skills. Another area of future research is to focus on the fundamental skills themselves. The Model of Expertise identifies 12 fundamental skills and including *collaborate* as we suggest here makes 13. Are these the only fundamental skills experts need to possess? Is it possible to identify additional fundamental skills?

### 3.5 Grammar of Action

We envision a cognitive systems future in which humans will interact with and collaborate with multiple cogs. Today, it is easy to imagine this interaction to be based on natural language, traditional computer displays, keyboards, mice, and touchscreens. However, we believe there will be and should be a greater bandwidth for communication between human and cog. An exciting area of future research is to explore these possibilities which might include: gesture recognition, body language, gait analysis, facial expressions, emotional awareness, activity recognition, and behavioral recognition among others.

Harper described the importance of considering the grammars of action of new technologies [49]. Whenever new technologies are adopted, human behavior changes to accommodate and integrate the new technology into our daily lives. As a result, a set of nouns, verbs, adjectives, and phrases arises because of the new technology. This is called a *grammar of action*.

For example, consider the computer mouse. Words like *click, double click, scroll wheel,* and *mouse pointer* describe how to perform tasks such as *select, drag, drop,* and *rubber band* items on the computer screen. These words existed before the computer mouse was invented but new meanings were adopted as a result of their use in the context of human/computer interaction. Similarly, *pinch, spread, tap, double tap, swipe left,* and *swipe right* are included in the grammar of action for touchscreen interfaces.

In the cog era, we expect a new grammar of action to arise as the result of mass-market adoption and use of cognitive systems. An interesting area of future research is to predict what new nouns, verbs, and adjectives will be invented. However, we expect more than words will be invented. What new gestures will be invented to facilitate easy and effective human/cog interaction? How will the cog use the human's facial expressions, body language, and emotional state to enhance and guide its interaction and collaboration with humans? We anticipate elements of the "cog grammar of action" to connect to the higher-order skills described in the last section. For example, one can imagine a certain hand gesture to become the normal way for a human to express to the cog "please explain that to me."

## 4   The Cog Era

We are at the very beginning of the cog era and its evolution will play out over the next few decades. The cog era will, for the first time, give humans artificial systems able to perform some of the high-level thinking. This will create a new industry and a new market sure to change things for us culturally, legally, socially, and legally.

*The Cog Market*
We expect the emergence of personal cogs intended for the mass-market. These cogs will be bought and sold by average people through existing Internet sales channels much in the same way apps, music, and other items are sold now. We will both be able to purchase or rent our own cogs and also be able to subscribe to cloud-based cog services and information. This will give every person access to professional-level expertise in any domain. This *democratization of expertise* will lead to changes similar in scope to the way the democratization of computing and information has changed us over the last few decades.

*Expertise as a Commodity*
Because cogs learn from humans, we expect the need for experts in a field to work with cogs and develop their own unique store of knowledge. Entities in industries such as financial services, investment services, legal, medical, news, politics, and technology will compete in offering access to their "superior" store of knowledge created through the interaction of their experts and their cogs. In the cog era, knowledge will become an economic commodity.

*Teacher Cogs*
We expect cogs to become intelligent tutoring systems. Through customized and personalized interaction with a person, teacher cogs will impart this knowledge to the student in ways similar to the master/apprentice model of education. The best teacher cogs will be personal cogs able to remember every interaction with a person over an extended period of time, even years or decades. Imagine an algebra cog able to answer a question by a 35-year old who it has been working with since grade school. We anticipate teacher cogs to evolve for every subject taught in schools and beyond. We think students of future generations will start using cogs all throughout their education and then retain the cogs, and years of interaction, through the rest of their lives. Again, we foresee vigorous competition arising from different teacher cog providers attempting to bring to the market the best teacher cog for a particular subject matter.

*Advisor, Coach, Self-Help, and Pet Cogs*
It is natural for humans to form emotional relationships with anything, biological or artificial, they can interact with. Indeed, people form emotional relationships with animals and technology today. We foresee cog technology giving personalities to artificial systems. Since cogs will be able to give expert-level advice in any domain, we predict the evolution of a host of self-help cogs ranging from relationship advice to life/work balance, grief counseling, faith-based counseling and beyond. People will confide intimate details to these cogs and receive advice of great personal value and satisfaction. People will spend hours conversing with their personal self-help/companionship cogs. We can easily envision the development of virtual pets with cog-based personalities and communication abilities. In the cog era, we will love our cog pets. Indeed, we have already witnessed the beginning of these kind of applications. Today, hundreds of millions of people have used "synthetic frend" cogs Xiaoice in China, Rinna in Japan, and Ruuh in India.

*Productivity Cogs*

We predict every productivity application in use today will become enhanced by cog technology in the future. Indeed, applications like word processors, spreadsheets, presentation editors, Web browsers, entertainment apps, games, graphics editing, etc. may become a primary interface point for humans and cogs. Cog capabilities will both be built into the applications themselves and provide expert-level collaboration to the user and also evolve into stand-alone cogs for a particular task. For example, we can imagine a future version of Microsoft Word coming complete with embedded creative writing cog services. We can also imagine purchasing a creative writing cog from an app store operating independently of a specific word processor.

Personal productivity cogs will understand our recent context in a deep manner and use that to customize their assistance and interaction with us. Imagine, for example, a word processing cog that understands you are writing about the future of cognitive processing but also knows that you have communicated with several others via email on that and related topics and can also take into consideration every article or Web page you have accessed in recent months while researching the paper. Such a cog knows a lot about you personally and can combine that knowledge with its own searching and reasoning about the millions of documents it has searched on the Internet. Personal productivity cogs will become our intelligent assistants.

*Collaborative Cognition*

In addition to enhancing current productivity applications, we expect an entirely new genre of cog-based productivity app to arise, *collaborative cognition*. We envision new kinds of problem solving, brainstorming, business/competitive/market analysis, and big data analysis. We foresee multi-cog "collaborative virtual team" applications being created. Collaborative cogs will become our artificial intelligent team members. Again, we see a vigorous and dynamic competitive market arising around the idea of collaborative cogs. By partnering with humans, cogs achieve ever-increasing levels of knowledge in a particular area. Therefore, considerable market value will be attached to collaborative cogs that have worked with the best experts in the field. The cog era will bring forth a new kind of virtual consultant.

*Research Cogs*

We foresee future graduate students, entrepreneurs, scientists and any of us creative and inquisitive people conducting research by conversing with their research cog(s) instead of searching and reading scores of journal articles and technical papers. Today, we tell graduate students the first step in their research is to go out and read as many articles, books, and papers as they can find about their topic. Future research students' first action will be to sit down with his research cog and ask "So, what is the current state of the art in <insert domain here>."

Cogs will far exceed the ability of any human in consuming billions of articles, papers, books, Web pages, emails, text messages, and videos. Even if a person spent their entire professional life learning and researching a particular subject is not able to read and understand everything available about that subject. Yet, future researchers will be able to start their education from that vantage point by the use of research cogs. In the cog era, the best new insights and discoveries will come from the interaction between researchers and their research cogs.

Here again we see evidence of knowledge becoming a commodity. Today, we may be able to learn a great deal from the notebooks of great inventors like Tesla, Edison, and DaVinci. In fact, notebooks of inventors like these are worth millions of dollars. But imagine how valuable it would be if we had access to Einstein's personal research cog he used for years while he was synthesizing the theory of relativity. In the cog era, not only will cogs assist us in coming up with great discoveries, they will also record and preserve that interaction for future generations. Such cogs will be enormously valuable both economically and socially.

*Discovery Engines*

Even though we envision cogs partnering with humans, we expect cogs to evolve to be able to perform on their own. We fully expect cogs working semi-autonomously to discover significant new theories, laws, proofs, associations, correlations, etc. In the cog era, the cumulative knowledge of the human race will increase by the combined effort of millions of cogs all over the world. In fact, we foresee an explosion of knowledge, an exponential growth, when cogs begin working with the knowledge generated by other cogs. This kind of cognitive work can proceed without the intervention of a human and therefore proceed at a dramatically accelerated rate. We can easily foresee the point in time where production of new knowledge by cogs exceeds, forever, the production of new knowledge by humans.

In fact, we anticipate a class of discovery engine cogs whose sole purpose is to reason about enormous stores of knowledge and continuously generate new knowledge of ever-increasing value resulting ultimately in new discoveries that would have never been discovered by humans or, at the very least, taken humans hundreds if not thousands of years to discover.

*Cognitive Property Rights*

The cog era will bring forth new questions, challenges, and opportunities in intellectual property rights. For example, if a discovery cog makes an important new discovery, who owns the intellectual property rights to that discovery? An easy answer might be "whoever owned the cog." But, as we have described, we anticipate cogs conferring with other cogs and using knowledge generated by other cogs. So a cog's work and results are far from being in isolation. We predict existing patent, copyright, trademark, and service mark laws will have to be extended to accommodate the explosion of knowledge in the cog era.

## 5  Democratization of Expertise

As described earlier, non-experts will be able to achieve or exceed expert-level performance in virtually any domain by working with cogs achieving Level 3 or Level 4 cognitive augmentation. When a large number of these cogs become available to the masses via the cog market as described above, we will be in a future where any average person could be a *synthetic expert*. We call this the *democratization of expertise.*

When expertise becomes available to the masses, changes will occur. Democratization of expertise will disrupt many professions. While we certainly do not anticipate the demise of doctors, lawyers, and accountants, their professions may change when their former customers have access to expert-level information and services possibly superior to what they could have supplied. What are the consequences and possibilities when millions of us can achieve expert-level performance in virtually any domain of discourse?

## 9  Conclusion

We have introduced the concept of *synthetic expertise* and have defined it as the ability of an average person to achieve expert-level performance by virtue of working with and collaborating with artificial entities (cogs) capable of high-level cognitive processing. Humans working in collaboration with cogs in a human/cog ensemble are cognitively augmented as a result of the collaboration. Over time, as the capabilities of cogs improve, humans will perform less and less of the thinking. We have used the balance of cognitive effort between human and cog to formulate the Levels of Cognitive Augmentation to describe the phenomenon. The Levels of Cognitive Augmentation can be used in the future to compare and contrast different systems and approaches.

To describe *expertise*, we have combined previous work in cognitive science, cognitive architectures, and artificial intelligence with the notion of expertise from education pedagogy to formulate the Model of Expertise. The Model of Expertise includes a Knowledge Level and an Expertise Level description of the fundamental knowledge and skills required by an expert.

Cogs will continue to improve and take on more of the skills defined in the Model of Expertise. These cogs will also become available to everyone via mass-market apps, services, and devices. Expertise becoming available to the masses is something we call the *democratization of expertise* and will usher in many social, cultural, societal, and legal changes.

We have identified several interesting areas of future research relating to synthetic expertise:

- Human/cog communication (HCI, AR, VR, ER, human/brain interfaces)
- Human/cog teaming and collaboration (HAT, CSCW)
- Cog/cog communication
- Fundamental skills of an expert
- Composite skills of an expert
- Grammar of action associated with cogs
- Human-centered design
- Task learning
- Problem-Solving method learning
- Goal assessment
- Strategic planning
- Common sense knowledge learning
- Intelligent agent theory
- Software agents
- Intellectual property
- Automated knowledge discovery

Some characteristics and challenges of the coming cognitive systems era have been described. Besides mass-market adoption of cog technology, we see expertise and knowledge becoming commodities leading us to interesting futures involving artificially-generated knowledge and legal battles over ownership of knowledge. We also describe a future in which people routinely collaborate with, learn from, and commiserate with cogs. In much the same way computer and Internet technology has woven itself into every fiber of life, we expect cognitive system technology to do the same.

# References

[1] Chapman, S. (1942). Blaise Pascal (1623-1662) Tercentenary of the calculating machine, *Nature,* London 150: 508–509.

[2] Hooper, R. (2015). Ada Lovelace: My brain is more than merely mortal, *New Scientist*, Internet page located at https://www.newscientist.com/article/dn22385-ada-lovelace-my-brain-is-more-than-merely-mortal last accessed November 2015.

[3] Isaacson, W. (2014). *The Innovators: How a Group of Hackers, Geniuses, and Geeks Created the Digital Revolution*, Simon & Schuster, New York, NY.

[4] Bush, V. (1945). As We May Think, *The Atlantic*, July.

[5] Ashby, W.R. (1956) *An Introduction to Cybernetics*, Chapman and Hall, London.

[6] [Apple] (1987). Knowledge Navigator, *You Tube* video located at: https://www.youtube.com/watch?v=JIE8xk6Rl1w Last accessed April 2016.

[7] Jackson, J. (2015). IBM Watson Vanquishes Human Jeopardy Foes, *PC World*. Internet page http://www.pcworld.com/article/219893/ibm_watson_vanquishes_human_jeopardy_foes.html last accessed May 2015.

[8] Ferrucci, D.A. (2012). Introduction to This is Watson*, IBM J. Res. & Dev.* Vol. 56 No. 3/4.

[9] Ferrucci, D., Brown, E., Chu-Carroll, J., Fan, J., Gondek, D., Kalyanpur, A., Lally, A., Murdock, J. W., Nyberg, E., Prager, J., Schlaefer, N., Welty, C. (2010). Building Watson: An Overview of the DeepQA Project, *AI Magazine*, Vol. 31, No. 3.

[10] Wladawsky-Berger, I. (2015). The Era of Augmented Cognition, *The Wall Street Journal: CIO Report*, Internet page located at http://blogs.wsj.com/cio/2013/06/28/the-era-of-augmented-cognition/ last accessed May.

[11] Gil, D. (2019). Cognitive systems and the future of expertise, *YouTube* video located at https://www.youtube.com/watch?v=0heqP8d6vtQ and last accessed May 2019.

[12] Kelly, J. E. and Hamm, S. (2013). *Smart Machines: IBMs Watson and the Era of Cognitive Computing*, Columbia Business School Publishing, Columbia University Press, New York, NY.

[13] Silver, D. et al., "Mastering the game of Go with deep neural networks and tree search", *Nature*, 529, 2016.

[14] [DeepMind] (2018a). The story of AlphaGo so far, *DeepMind* Internet page: https://deepmind.com/research/alphago/ last accessed February 2018.

[15] [DeepMind] (2018b). AlphaGo Zero: learning from scratch, *DeepMind* Internet page: https://deepmind.com/blog/alphago-zero-learning-scratch/ last accessed February 2018.

[16] [ChessBase] (2018). AlphaZero: Comparing Orangutans and Apples, *ChessBase* Internet page: https://en.chessbase.com/post/alpha-zero-comparing-orang-utans-and-apples last accessed February 2018.

[17] Licklider, J.C.R. (1960). Man-Computer Symbiosis, *IRE Transactions on Human Factors in Electronics*, Vol. HFE-1, March.

[18] Engelbart, D. C. (1962). Augmenting Human Intellect: A Conceptual Framework, *Summary Report AFOSR-3233*, Stanford Research Institute, Menlo Park, CA, October.

[19] Newell, A. (1990). *Unified Theories of Cognition,* Harvard University Press, Cambridge, Massachusetts.

[20] Newell, A. (1982) The Knowledge Level, *Artificial Intelligence*, 18(1):87-127.

[21] Chase, W. and Simon, H. (1973). Perception in Chess, *Cognitive Psychology* Volume 4.

[22] Steels, L. (1990). Components of Expertise, *AI Magazine*, Vol 11, No. 2.

[23] Genesereth, M. and Nilsson, N. (1987). *Logical Foundations of Artificial Intelligence,* Morgan Kaufmann.

[24] Russell, S. and Norvig, P. (2009). *Artificial Intelligence: A Modern Approach, 3rd Edition,* Pearson.

[25] Laird, J. E. (2012). *The SOAR Cognitive Architecture,* The MIT Press.

[26] Bloom, B. S., Engelhart, M. D., Furst, E. J., Hill, W. J., Krathwohl, D. R. (1956). *Taxonomy of educational objectives: The classification of educational goals. Handbook I: Cognitive domain*. New York: David McKay Company.

[27] Anderson, L. W., and Krathwohl, D. R. (eds) (2001). *A taxonomy for learning, teaching, and assessing: A revision of Bloom's taxonomy of educational objectives*.

[28] Fulbright, R. (2017). Cognitive Augmentation Metrics Using Representational Information Theory, in: Schmorrow D., Fidopiastis C. (eds) *Augmented Cognition. Enhancing Cognition and Behavior in Complex Human Environments*. AC 2017. Lecture Notes in Computer Science, vol 10285. Proceedings of HCI International 2017 conference, Springer.

[29] Fulbright, R. (2018). On Measuring Cognition and Cognitive Augmentation, in: Yamamoto, S. and Mori, H. (eds) *Human Interface and the Management of Information*. LNCS 10904. Proceedings of HCI International 2018 conference, Springer.

[30] Fulbright, R. (2019). Calculating Cognitive Augmentation –A Case Study, *Proceedings of HCI International 2019* conference, Springer.

[31] Fulbright, R. (2016). How Personal Cognitive Augmentation Will Lead To The Democratization of Expertise, *Fourth Annual Conference on Advances in Cognitive Systems*, Evanston, IL, June. Available online at: http://www.cogsys.org/posters/2016, last retrieved January 2017.

[32] Fulbright, R. (2016). The Cogs Are Coming: The Coming Revolution of Cognitive Computing, *Proceedings of the 2016 Association of Small Computer Users in Education* (ASCUE) Conference, June.

[33] Fulbright, R. (2017). ASCUE 2067: How We Will Attend Posthumously, *Proceedings of the 2017 Association of Small Computer Users in Education* (ASCUE) Conference, June.

[34] Gobet, F. (2016). *Understanding Expertise: A Multidisciplinary Approach*, Palgrave, UK.

[35] Gobet, F. and Simon, H. (2000). Five Seconds or Sixty? Presentation Time in Expert Memory, *Cognitive Science,* Vol. 24 (4).

[36] Gobet, F. and Chassy, P. (2009). Expertise and Intuition, A Tale of Three Theories, *Minds & Machines*, Springer, 2009.

[37] Dreyfus, H. L. (1972). *What Computers Can't Do: A Crtique of Artificial Reason*, The MIT Press, Cambridge, MA.

[38] Dreyfus, H. L. and Dreyfus, S. E. (1988). *Mind Over Machine: The Power of Human Intuition and Expertise in the Era of the Computer,* New York: Free Press.

[39] Minsky, M. (1977). Frame-System Theory, in P.N. Johnson-Laird & P.C. Watson (eds), *Thinking. Readings in Cognitive Science*, Cambridge University Press.

[40] Wehner, M. (2019). AI is now better at predicting mortality than human doctors, *New York Post,* published May 14, 2019. Available online at: https://nypost.com/2019/05/14/ai-is-now-better-at-predicting-mortality-than-human-doctors/?utm_campaign=partnerfeed&utm_medium=syndicated&utm_source=flipboard, and last accessed June, 2019.

[41] Lavars, N. (2019). Machine learning algorithm detects signals of child depression through speech, *New Atlas,* published May 7, 2019, Available online at: https://newatlas.com/machine-learning-algorithm-depression/59573/ and last accessed June 2019.

[42] Sandoiu, A. (2019). Artificial intelligence better than humans at spotting lung cancer, *Medical News Today Newsletter*, May 20, Available online at: https://www.medicalnewstoday.com/articles/325223.php#1, last accessed November 2019.

[43] Towers-Clark, C. (2019). The Cutting-Edge of AI Cancer Detection, *Forbes,* published April 30, 2019, Available online at: https://www.forbes.com/sites/charlestowersclark/2019/04/30/the-cutting-edge-of-ai-cancer-detection/#45235ee77336 and last accessed June 2019.

[44] Gregory, M., (2019). AI Trained on Old Scientific Papers Makes Discoveries Humans Missed, *Vice,* Available online at: https://www.vice.com/en_in/article/neagpb/ai-trained-on-old-

[45] De Groot, A. D. (1965). *Thought and Choice in Chess*, The Hague, Mouton.

[46] Fulbright, R. (2020). The Expertise Level, *Proceedings of HCI International 2020* conference, Springer.

[47] Haenssle, H. A., Fink, C., Schneiderbauer, R., Toberer, F., Buhl, T., Blum, A., Kalloo, A., Hassen, A. B. H., Thomas, L., Enk, A. and Uhlmann, L. (2018). Man against machine: diagnostic performance of a deep learning convolutional neural network for dermoscopic melanoma recognition in comparison to 58 dermatologists, *Annals of Oncology*, Vol. 29, Issue 8, August, Pages 1836–1842. Available online at: https://academic.oup.com/annonc/article/29/8/1836/5004443, last accessed November 2019.

[48] Abràmoff, M. D., Lavin, P. T., Birch, M., Shah, N. and Folk, J. C. (2018). Pivotal trial of an autonomous AI-based diagnostic system for detection of diabetic retinopathy in primary care offices, *Digital Med* Vol. 1, 39.

[49] Harper, R. (2019). The Role of HCI in the Age of AI, *International Journal of Human–Computer Interaction*. Volume 35. Available online at: https://www.tandfonline.com/doi/abs/10.1080/10447318.2019.1631527 and last accessed January 2020.